\documentclass[twocolumn,showpacs,preprintnumbers,amsmath,amssymb]{revtex4}

\usepackage{graphicx}
\usepackage{dcolumn}
\usepackage{bm}

\begin{document}

\preprint{APS/123-QED}

\title{Impurity effects of $\Lambda$ particle on the 2$\alpha$ cluster states of $^{9}$Be and $^{10}$Be}

\author{M. Isaka$^1$ and M. Kimura$^2$}
\affiliation{$^1$RIKEN Nishina Center, RIKEN, Wako, Saitama 351-0198,Japan}
\affiliation{$^2$Department of Physics, Hokkaido University, Sapporo 060-0810,Japan}

%

\date{\today}

\begin{abstract}
The low-lying structure of $^{10}_\Lambda$Be and $^{11}_\Lambda$Be are investigated within the framework of the antisymmetrized molecular dynamics. 
We focus on the modifications of the excitation spectra and dynamical changes of the 2$\alpha$ cluster structure caused by a $\Lambda$ particle as an impurity in these hypernuclei. 
It is found that the excitation energies of well-pronounced cluster states are largely shifted up by the addition of a $\Lambda$ particle. Furthermore, we also find that the 2$\alpha$ cluster structure is significantly changed in the excited states, whereas it is almost unchanged in the ground states.
\end{abstract}

\pacs{Valid PACS appear here}
\maketitle

\section{Introduction}

Structure of hypernuclei has been intensively investigated to study hyperon-nucleon and hyperon-hyperon interactions \cite{PRC31.499(1985), PPNP57.564(2006), NPA804.84(2008), PPNP63.339(2009)} and to understand dynamics of baryon many-body systems. 
For the latter purpose, many theoretical works have been carried out in $p$-$sd$-$pf$ shell $\Lambda$ hypernuclei and revealed structure changes caused by the addition of a $\Lambda$ particle, so-called \textit{impurity effects}, such as changes of cluster structures \cite{PTP70.189(1983),PTP71.985(1984),PTPS81.42(1985),PTP78.1317(1987),PRC59.2351(1999),PRL85.270(2000),PRC83.054304(2011),NPA881.288(2012),PTP128.105(2012),PRC91.014314(2015)} and deformation \cite{PRC76.034312(2007),PRC78.054311(2008),PTP123.569(2010),PRC83.044323(2011),PRC84.014328(2011),PRC90.064302(2014),PRC91.024327(2015),PRC91.054306(2015)}. 
In particular, $p$-shell nuclei with pronounced clustering manifest drastic structure changes. 
For example, it is well known that a $\Lambda$ particle reduces the inter-cluster distance between $\alpha$ and $d$ clusters in $^7_\Lambda$Li \cite{PTPS81.42(1985), PRC59.2351(1999), PRL86.1982(2001)}. 
An analogous reduction of the inter-cluster distance was also predicted for $\alpha$ and $^{16}$O clusters in $^{21}_\Lambda$Ne \cite{PTP71.985(1984), PRC83.054304(2011)}. 
In $^{13}_\Lambda$C, it is discussed that the $\Lambda$ particle significantly reduces the size of the Hoyle state $^{12}{\rm C}(0^+_2)$ having a dilute 3$\alpha$ cluster structure \cite{PTP97.881(1997)}. 
In addition to the dynamical changes of cluster structure, it is also predicted that the $\Lambda$ binding energy $B_\Lambda$ of the $^{12}{\rm C}(0^+_2) \otimes \Lambda$ state is smaller than those of the compact shell-model like states by about 3 MeV \cite{PRL85.270(2000)}. 
This is because the dilute 3$\alpha$ cluster structure of the Hoyle state makes the attraction between $\Lambda$ and nucleons weaker compared to the compact ground state.

From the above point of view, structure of Be hyper-isotopes is particularly of interest, since Be isotopes have 2$\alpha$ cluster core surrounded by valence neutrons \cite{PTP57.866(1977),PTP59.315(1978),PTP61.1049(1979),PTP65.204(1981),PRC61.044306(1999),PTEP01A.201(2012)}. 
For example, in $^9$Be, the first excited state $1/2^+$ is considered to have $^{8}$Be($0^+$) + $n(s_{1/2})$ configuration, which can be regarded as a Hoyle analogue state with the replacement of a $\alpha$ particle by a neutron, while the ground state has a relatively compact structure with $^{8}$Be($0^+$) + $n(p_{3/2})$ configuration. 
It is known that the $1/2^+$ state of $^9$Be is of significance in nuclear astrophysics, because this state has an impact on the reaction rate of $^8{\rm Be} (n, \gamma) ^{9}{\rm Be}$ in stellar environments and supernova explosions, which plays an quite important role in nucleosynthesis processes \cite{AstroJ562.470(2001),AstroJ634.1173(2005)}. 
Owing to its importance, the nature of the $1/2^+$ state has been investigated with the $\alpha$ + $\alpha$ + $n$ models by many authors, but the conclusions are still controversial. 
In Refs. \cite{EPJA4.33(1999),PRC68.014310(2003)}, the virtual-state character of this state was shown, while the authors of Refs. \cite{PLB684.132(2010),PRC82.034001(2010)} discussed this state as a three-body resonance state. 
It is expected that study of $^{10}_\Lambda$Be provides a new insight to the nature of the $1/2^+$ state, because a $\Lambda$ particle will bound this state and spectroscopic information could be obtained. 
In neutron-rich side, exotic structures associated with the 2$\alpha$ clustering appear. 
For example, in $^{11}$Be, it is well known that the ground-state parity is positive ($1/2^+$), whereas it is expected to be $1/2^-$ in ordinary shell model picture \cite{PR113.563(1959),PRL4.469(1960),NPA248.1(1975)}. This is referred as \textit{parity-inverted ground state}, and explained in terms of the molecular orbits of valence neutrons around the 2$\alpha$ clusters \cite{PRC66.024305(2002)}. 
In our previous work \cite{PRC91.014314(2015)}, we predicted that the parity-inverted ground state of $^{11}$Be is reverted by the addition of a $\Lambda$ particle. 
Specifically, the $^{11}{\rm Be}(1/2^-) \otimes \Lambda$ state becomes the ground state, because the $^{11}{\rm Be}(1/2^-) \otimes \Lambda$ configuration has larger $\Lambda$ binding energy $B_\Lambda$ than the $^{11}{\rm Be}(1/2^+) \otimes \Lambda$ configuration reflecting their different structures. 
The coexistence of the different structures has been also discussed in $^{10}$Be. In this nucleus, the two valence neutrons are considered to occupy different orbits in the $0^+_1$, $0^+_2$ and $1^-$ states \cite{PRC60.064304(1999)}, and depending on the neutron occupation, the degree-of-clustering varies. 
In particular, the $0^+_2$ state is considered to be largely deformed having a well-developed 2$\alpha$ cluster structure. Therefore, we can expect the modification of the excitation spectra by the addition of a $\Lambda$ particle. 
Furthermore, it is also interesting to investigate the dynamical changes of these structures and compare it with those in $^{10}_\Lambda$Be and $^{12}_\Lambda$Be.

The aim of the present study is to reveal the modifications of the excitation spectra as well as the dynamical changes of the 2$\alpha$ cluster structure by the addition of a $\Lambda$ particle into $^{9}$Be and $^{10}$Be. 
Focusing on the ground and excited states, the difference of $\Lambda$ binding energies $B_\Lambda$ is investigated with the framework of the antisymmetrized molecular dynamics for hypernuclei (HyperAMD) \cite{PRC83.044323(2011)}. 
It is found that the $B_\Lambda$ in the excited states with well-developed cluster structure are much smaller than those in the compact ground states in $^{10}_\Lambda$Be and $^{11}_\Lambda$Be. We also show that the $\Lambda$ particle significantly reduces the inter-cluster distance between the 2$\alpha$ clusters of these excited states. 

This paper is organized as follows. In the next section, the theoretical framework of HyperAMD is explained. 
In Sec. III, the modifications of the excitation spectra associated with the difference of $B_\Lambda$ and dynamical changes of the cluster structure are discussed. The final section summarizes this work.

\section{Framework}

In this study, we apply the HyperAMD combined with the generator coordinate method (GCM) \cite{PRC83.054304(2011)} to $^{10}_\Lambda$Be and $^{11}_\Lambda$Be hypernuclei. 

\subsection{Hamiltonian and variational wave function}

The Hamiltonian used in this study is given as,
\begin{eqnarray}
\hat{H} &=& \hat{H}_\Lambda + \hat{H}_N - \hat{T}_g,\\
\hat{H}_\Lambda &=& \hat{T}_{\Lambda} + \hat{V}_{\Lambda N},\\
\hat{H}_N &=& \hat{T}_{N} + \hat{V}_{NN} + \hat{V}_{Coul}.
\label{HN}
\end{eqnarray}
Here, $\hat{T}_{N}$, $\hat{T}_{\Lambda}$ and $\hat{T}_g$ are the kinetic energies of nucleons, a $\Lambda$ particle and the center-of-mass motion, respectively.
We use the Gogny D1S interaction \cite{PRC21.1568(1980)} as an effective nucleon-nucleon interaction $\hat{V}_{NN}$. The Coulomb interaction $\hat{V}_{Coul}$ is approximated by the sum of seven Gaussians. 
As the $\Lambda N$ effective interaction $\hat{V}_{\Lambda N}$, we use the same $YN$ $G$-matrix interactions as in our previous work for $^{12}_\Lambda$Be \cite{PRC91.014314(2015)}, derived from the Nijmegen potentials named model D \cite{PRD15.2547(1977),NPA547.245(1992)}, NSC97f \cite{PRC59.21(1999)} and ESC08c \cite{PTPS185.14(2010),Gen57.6(2013)}, which we call ND, NSC97f and ESC08c, respectively. 
As the spin-orbit interaction part of $\hat{V}_{\Lambda N}$, we always use that of ESC08c.

The intrinsic wave function of a single $\Lambda$ hypernucleus composed of a core nucleus with mass number $A$ and a $\Lambda$ particle is described by the parity-projected wave function, $\Psi^\pi = \hat{P}^\pi \Psi_{int}$, where $\hat{P}^\pi$ is the parity projector and $\Psi_{int}$ is the intrinsic wave function given as, 
\begin{eqnarray}
\Psi_{int} &=& \Psi_N \otimes \varphi_\Lambda,\quad
\Psi_N = \frac{1}{\sqrt{A!}}\det \left\{ \phi_{i} \left( r_j \right) \right\},\\
\phi_{i} &=& \prod_{\sigma=x,y,z} \biggl(\frac{2\nu_\sigma}{\pi}\biggr)^{\frac{1}{4}}
 e^{-\nu_\sigma \bigl(r - Z_{i} \bigr)_\sigma^2 } \chi_{i} \eta_{i} ,\\
\varphi_\Lambda &=& \sum_{m=1}^M c_m \chi_m \prod_{\sigma=x,y,z} \biggl(\frac{2\nu_\sigma}{\pi}\biggr)^{\frac{1}{4}}
 e^{-\nu_\sigma \bigl(r - z_m \bigr)_\sigma^2},
\end{eqnarray}
\begin{eqnarray}
\chi_{i} &=& \alpha_i \chi_\uparrow + \beta_i \chi_\downarrow,\quad \chi_m = a_m \chi_\uparrow + b_m \chi_\downarrow, \\
\eta_{i} &=& {\rm proton}\ {\rm or}\ {\rm neutron},
\end{eqnarray}
where $\phi_{i}$ is {\it i}th nucleon single-particle wave packet consisting of spatial, spin $\chi_{i}$ and isospin $\eta _{i}$ parts. 
The variational parameters are the centroids of Gaussian $\bm{Z}_i$ and $\bm{z}_m$, width parameters $\nu_\sigma$, spin directions $\alpha_i$, $\beta_i$, $\alpha_m$ and $\beta_m$, and coefficients $c_m$. We approximately remove the spurious center-of-mass kinetic energy in the same way as Ref. \cite{PRC83.044323(2011)}.

In the actual calculation, the energy variation is performed under the constraint on nuclear quadrupole deformation parameter $\beta$ in the same way as our previous works \cite{PRC83.044323(2011),PRC91.014314(2015)}. 
By the frictional cooling method, the variational parameters in $\Psi^\pi$ are determined for each $\beta$, and the resulting wave functions are denoted as $\Psi^\pi (\beta)$. 
It is noted that the nuclear quadrupole deformation parameter $\gamma$ is optimized through the energy variation for each $\beta$.
It is found that the $\Lambda$ particle dominantly occupies an $s$ orbit in the hypernuclei, because no constraint is imposed on the $\Lambda$ single-particle wave function in the present calculation. 

\subsection{Angular momentum projection and GCM}

After the variational calculation, we project out an eigenstate of the total angular momentum $J$ from the hypernuclear states, 
\begin{eqnarray}
 \label{AngProj}
 \Psi^{J\pi}_{MK}(\beta) &=& \frac{2J+1}{8\pi^2} \int d\Omega D^{J*}_{MK}(\Omega) \hat{R}(\Omega) \Psi^{\pi} (\beta).
\end{eqnarray}
The integrals are performed numerically over three Euler angles $\Omega$.

The wave functions $\Psi^{J \pi}_{MK} (\beta)$ which have the same parity and angular momentum but have different $K$ and nuclear quadrupole deformation $\beta$ are superposed (GCM). 
Then the wave function of the system is written as
\begin{eqnarray}
\label{GCM_wf}
\Psi^{J\pi}_{\alpha} &=& c_\alpha \Psi^{J \pi}_{MK}(\beta) + c'_\alpha \Psi^{J \pi}_{MK'}(\beta') + \cdots ,
\end{eqnarray}
where the quantum numbers except for the total angular momentum and the parity are represented by $\alpha$. The coefficients $c_\alpha$, $c'_\alpha$, $\cdots$ are determined by the Hill-Wheeler equation.

\subsection{Analysis of wave function}

The $\Lambda$ binding energy $B_\Lambda$ is calculated with the energies obtained by the GCM calculation. 
Namely, $B_\Lambda$ is defined as the energy gain of a $J^\pi$ state in a hypernucleus $^{A+1}_{\ \ \ \Lambda}$Be from the core state $j^\pi$ in $^{A}$Be, as, 
\begin{eqnarray}
B_\Lambda = E(^{A}{\rm Be}; j^\pi) - E(^{A+1}_{\ \ \ \Lambda}{\rm Be}; J^\pi).
\label{BLmd}
\end{eqnarray}
Here, $E(^{A}{\rm Be}; j^\pi)$ and $E(^{A+1}_{\ \ \ \Lambda}{\rm Be}; J^\pi)$ respectively represent the total energies of the $j^\pi$ state of the core nucleus and the corresponding $J^\pi$ state of the hypernucleus.
To investigate the difference of $B_\Lambda$, we calculate the expectation values of $\hat{T}_\Lambda$ ($T_\Lambda$) and $\hat{V}_{\Lambda N}$ ($V_{\Lambda N}$), and the energy of the nuclear part $E_N$ as, 
\begin{eqnarray}
E_N (^{A+1}_{\ \ \ \Lambda}{\rm Be}; J^\pi) = \langle \Psi^{J \pi}_{\alpha} | \hat{H}_N | \Psi^{J \pi}_{\alpha} \rangle,
\label{EN}
\end{eqnarray} 
where, $\hat{H}_N$ is defined by Eq. (\ref{HN}).

We introduce the overlap between the $\Psi^{J \pi}_{MK} (\beta)$ and GCM wave function $\Psi^{J\pi}_\alpha$, 
\begin{eqnarray}
O^{J\pi}_{MK\alpha} ( \beta ) = | \langle \Psi^{J \pi}_{MK} ( \beta ) | \Psi^{J \pi}_\alpha \rangle |^2,
\label{Overlap}
\end{eqnarray}
which we call GCM overlap.
Since $ O^{J\pi}_{MK \alpha} ( \beta )$ shows the contributions from $\Psi^{J \pi}_{MK} ( \beta )$ to each state $J^\pi$, it is useful to estimate the nuclear quadrupole deformation $\beta$ of each state. 
Namely, we regard $\beta$ corresponding to the maximum GCM overlap as the nuclear deformation of each state.

\begin{figure*}
  \begin{center}
    \includegraphics[keepaspectratio=true,width=172mm]{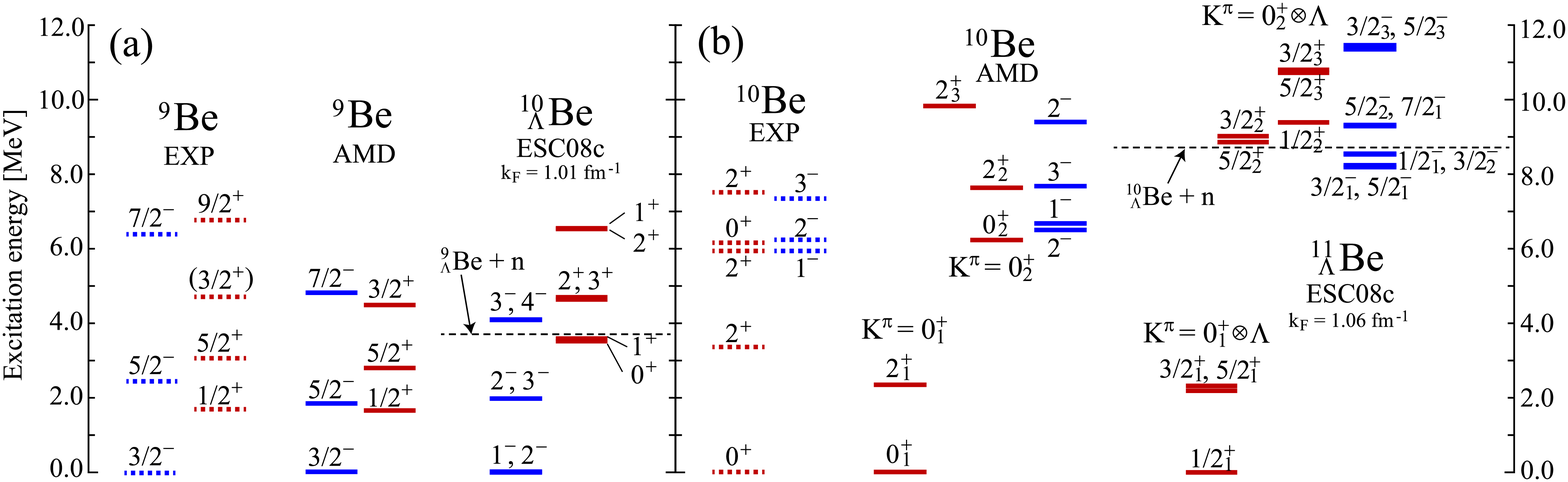}
  \end{center}
  \caption{(Color online) (a) Calculated excitation spectra of $^{9}$Be and $^{10}_\Lambda$Be. For the comparison, observed spectrum of $^{9}$Be \cite{NPA745.155(2004)} is also presented. Dashed line shows the experimental lowest threshold of $^{9}_\Lambda$Be + $n$ in $^{10}_\Lambda$Be. (b) Same as (a) but for $^{10}$Be and $^{11}_\Lambda$Be. 
  In $^{11}_\Lambda$Be, the energy of the lowest threshold $^{10}_\Lambda$Be + $n$ is estimated from the observed data of $^{10}_\Lambda$Be \cite{Gogami} and the empirical value of $B_\Lambda$ \cite{ANP8.1(1975)} due to the absence of the observation of $^{11}_\Lambda$Be.
}
  \label{fig:Ex.eps}
\end{figure*}

\begin{figure*}
  \begin{center}
    \includegraphics[keepaspectratio=true,width=160mm]{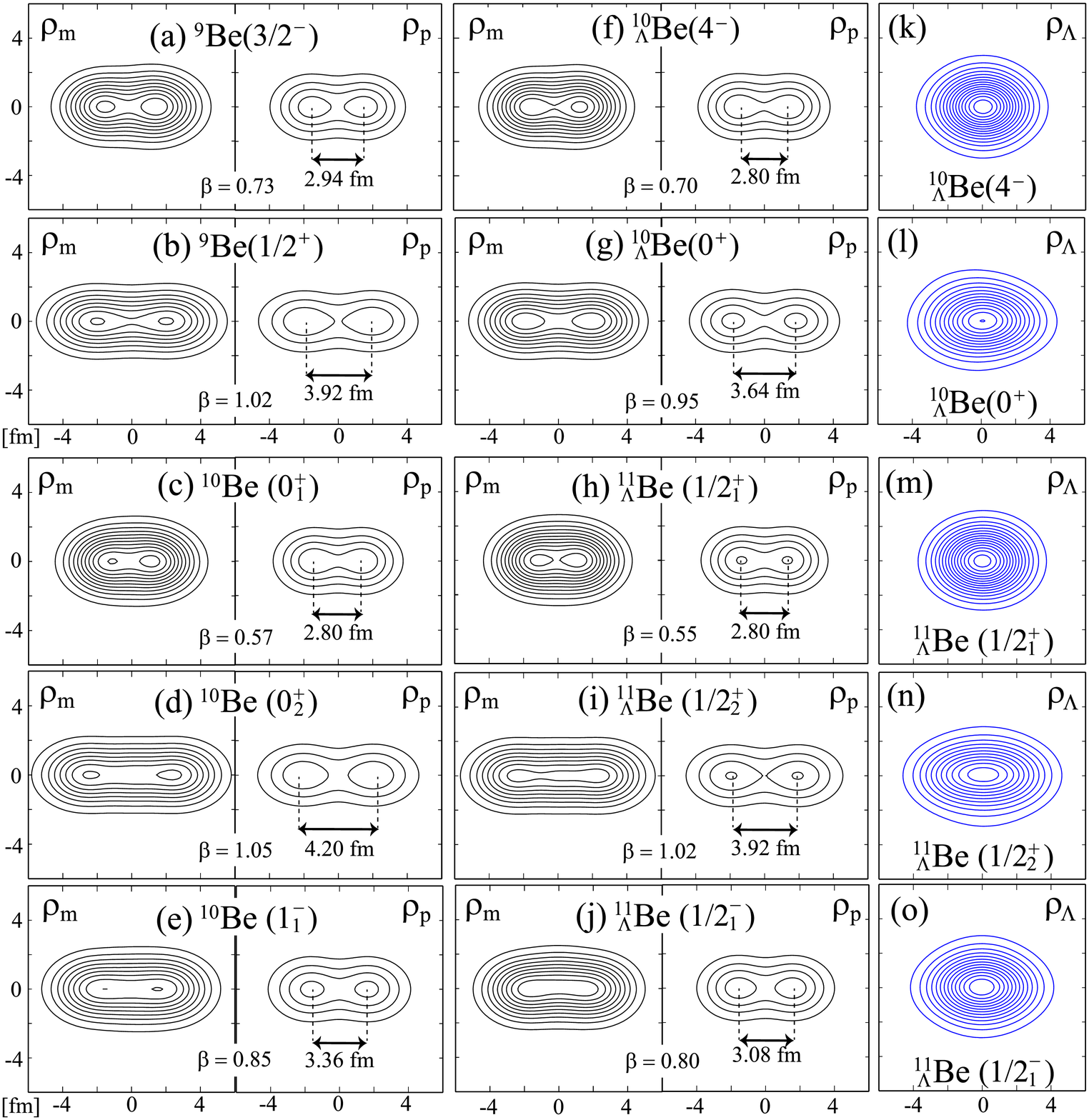}
  \end{center}
  \caption{(Color online) (a) - (e): Nuclear ($\rho_m$) and proton ($\rho_p$) density distributions of the $3/2^-$ and $1/2^+$ states in $^{9}$Be, and the $0^+_1$, $0^+_2$ and $1^-$ in $^{10}$Be. 
Inter-cluster distance between 2$\alpha$ estimated by the peaks of the proton density is also displayed in each state. 
(f) - (j): Nuclear and proton density distributions of the hypernuclei corresponding to (a) - (e). 
(k) - (o): Density distributions of the $\Lambda$ particle in each state in the hypernuclei. 
}
  \label{fig:dens_i.eps}
\end{figure*}

\section{Results and Discussions}
\label{sec3}

\begin{figure*}
  \begin{center}
    \includegraphics[keepaspectratio=true,width=172mm]{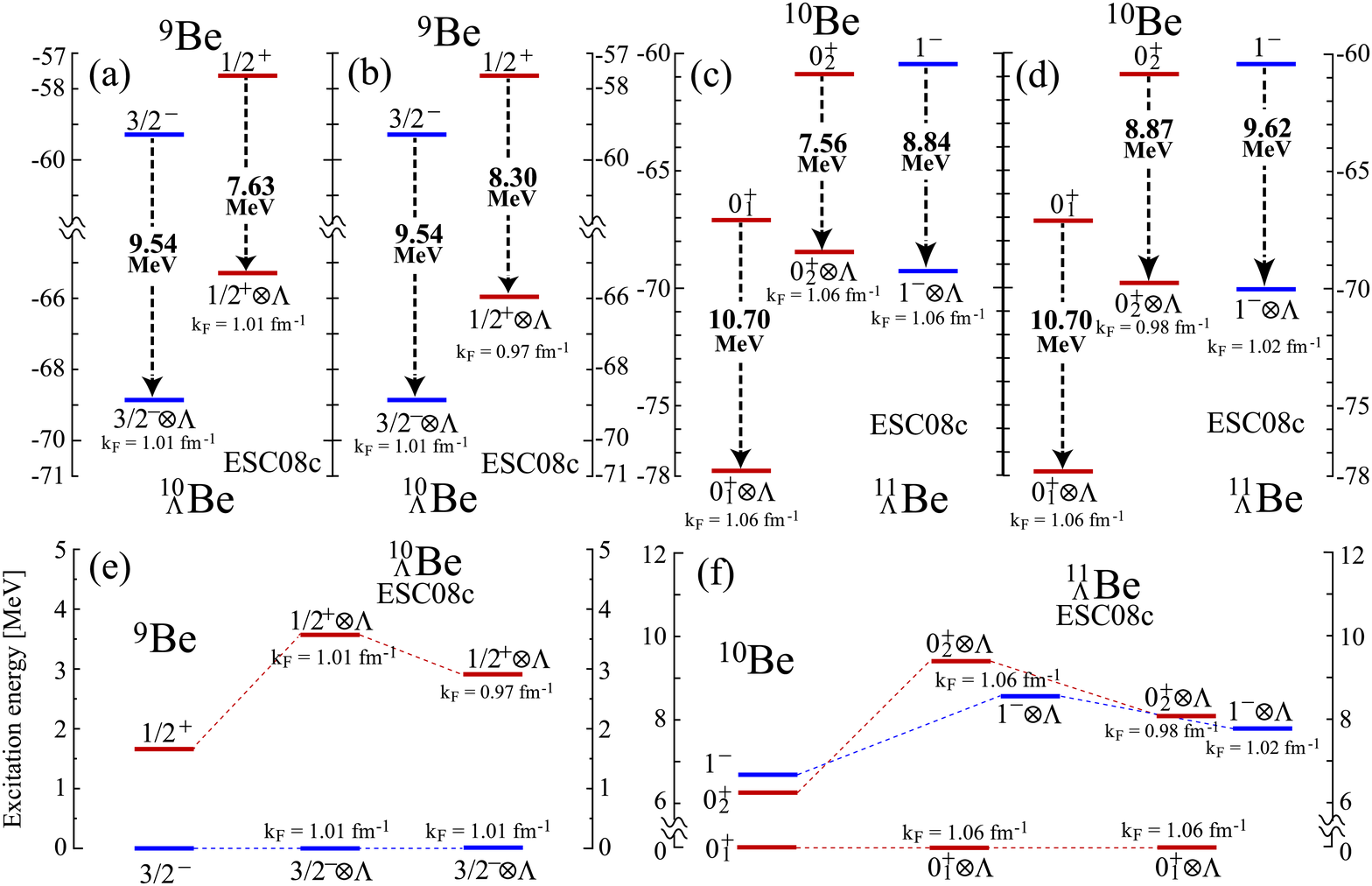}
  \end{center}
  \caption{(Color online) 
$\Lambda$ biding energies $B_\Lambda$ in the $^9{\rm Be}(3/2^-) \otimes \Lambda$ and $^9{\rm Be}(1/2^+) \otimes \Lambda$ states in $^{10}_\Lambda$Be with (a) the same  and (b) different $k_F$ values. 
Details of the $k_F$ treatment are explained in text. 
(c) and (d): Same as (a) and (b), respectively, but for $^{11}_\Lambda$Be. 
 (e) Comparison of the calculated excitation energies of the $^{9}{\rm Be}(1/2^+)$ state (left) and the corresponding state in $^{10}_\Lambda$Be with the $k_F = 1.01$ fm$^{-1}$ (middle) and $k_F = 0.97$ fm$^{-1}$ (right). Excitation energies are calculated with the centroid energies of the doublet states. 
(f) Same as (e), but for $^{11}_\Lambda$Be.
}
  \label{fig:gain-ESC.eps}
\end{figure*}

\begin{table*}
  \caption{Calculated total ($E$) and excitation ($E_x$) energies in MeV, matter quadrupole deformation $\beta$ of the $3/2^-$ and $1/2^+$ states in $^9$Be and the corresponding states in $^{10}_\Lambda$Be. 
  In $^{10}_\Lambda$Be, $\Lambda$ biding energies $B_\Lambda$ defined by Eq. (\ref{BLmd}), the expectation values of the $\Lambda$ kinetic ($T$) and $\Lambda N$ potential ($V_{\Lambda N}$), and the energy of the nuclear part $E_N$ defined by Eq. (\ref{EN}) are also listed in unit of MeV. 
Numbers in parentheses are observed values of $E$ and $E_x$ \cite{NPA729.337(2003),NPA745.155(2004)}, and $B_\Lambda$ \cite{NPA547.369(1992),Gogami}.
}
  \label{Tab:table1}
  \begin{ruledtabular}
  \begin{tabular}{cccccccccccc}
  \multicolumn{8}{c}{$^{10}_\Lambda$Be} & \multicolumn{4}{c}{$^{9}$Be}\\
  \cline{1-8} \cline{9-12}
  $J^\pi$ & $E$ & $E_x$ & $\beta$ & $E_N$ & $B_\Lambda$ & $T_\Lambda$ & $V_{\Lambda N}$ & $J^\pi$ & $E$ & $E_x$ & $\beta$ \\
  \hline\hline
  $2^-$ & -68.87 & 0.00 & 0.66 & -59.13 & 9.54 & 7.14 & -16.87 & $3/2^-$ & -59.32 & 0.00 & 0.73 \\
  $1^-$ & -68.85 & 0.02 & 0.70 & -59.11 & 9.53 & 7.23 & -16.96 &  & (-58.16) & (0.00) &  \\
  $^{9}{\rm Be}(3/2^-) \otimes \Lambda$ & -68.86 & 0.00 &  &  & 9.54 &  &  &  &  &  &  \\
        &        &  &  &  & ($9.11 \pm 0.22$ \cite{NPA547.369(1992)}) &  &  &  &  &  &  \\
        &        &  &  &  & ($8.55 \pm 0.18$ \cite{Gogami}) &  &  &  &  &  &  \\
  \hline
  $0^+$ & -65.35 & 3.52 & 1.00 & -57.34 & 7.68 & 6.44 & -14.45 & $1/2^+$ & -57.66 & 1.66 & 1.02 \\
  $1^+$ & -65.27 & 3.59 & 1.00 & -57.35 & 7.61 & 6.49 & -14.42 &  & (-56.48) & (1.68) &  \\
  $^{9}{\rm Be}(1/2^+) \otimes \Lambda$ & -65.29 & 3.57 &  &  & 7.63 &  &  &  &  &  & \\
  \end{tabular}
  \end{ruledtabular}
\end{table*}

\begin{table}
  \caption{Calculated proton ($r_p$), neutron ($r_n$) and matter ($r_m$) r.m.s. radii (fm) in the $3/2^-$ and $1/2^+$ states of $^9$Be and the corresponding states of $^{10}_\Lambda$Be.  
Numbers in parenthesis are observed point proton radii determined from the r.m.s. charge radii \cite{PRL102.062503(2009)} in the same way as in Ref. \cite{PRC91.014315(2015)}.
  }
  \label{Tab:table3}
  \begin{ruledtabular}
  \begin{tabular}{cccccccc}
  \multicolumn{4}{c}{$^{9}$Be} & \multicolumn{4}{c}{$^{10}_\Lambda$Be}\\
  \cline{1-4}\cline{5-8}
  $J^\pi$ & $r_p$ & $r_n$ & $r_m$ & $J^\pi$ & $r_p$ & $r_n$ & $r_m$ \\
  \hline\hline
  $3/2^-$ & 2.51 & 2.54 & 2.53 & $1^-$ & 2.43 & 2.47 & 2.45 \\
          &(2.38) &      &      & $2^-$ & 2.43 & 2.47 & 2.45 \\
  $1/2^+$ & 2.83 & 3.01 & 2.93 & $0^+$ & 2.71 & 2.91 & 2.83 \\
          &      &      &      & $1^+$ & 2.70 & 2.90 & 2.82 \\
  \end{tabular}
  \end{ruledtabular}
\end{table}

\begin{table*}
  \caption{Same as Tab.\ref{Tab:table1} but for $^{10}$Be and $^{11}$Be. Observed values are taken from Refs. \cite{NPA729.337(2003),NPA745.155(2004)}.
  }
  \label{Tab:table2}
  \begin{ruledtabular}
  \begin{tabular}{cccccccccccc}
  \multicolumn{8}{c}{$^{11}_\Lambda$Be} & \multicolumn{4}{c}{$^{10}$Be}\\
  \cline{1-8} \cline{9-12}
  $J^\pi$ & $E$ & $E_x$ & $\beta$ & $E_N$ & $B_\Lambda$ & $T_\Lambda$ & $V_{\Lambda N}$ & $J^\pi$ & $E$ & $E_x$ & $\beta$ \\
  \hline\hline
  $1/2^+_1$ & -77.82 & 0.00  & 0.498 & -67.16 & 10.70 & 7.30 & -17.96 & $0^+_1$ & -67.11 & 0.00 & 0.55 \\
   &  &  &  &  &  &  &  &  & (-64.98) & (0.00) &  \\
  \hline
  $1/2^+_2$ & -68.43 & 9.39 & 1.00 & -60.62 & 7.56 & 6.39 & -14.20 & $0^+_2$ & -60.87 & 6.24 & 1.05 \\
   &  &&  &  &  &  &  &  & (-58.80)  & (6.18) & \\
  \hline
  $1/2^-$ & -69.28 & 8.53 & 0.80 & -60.27 & 8.85 & 6.72 & -15.73 & $1^-$ & -60.43 & 6.68 & 0.85 \\
  $3/2^-$ & -69.27 & 8.55 & 0.77 & -60.22 & 8.84 & 6.72 & -15.77 &  & (-59.02) & (5.96) &  \\
  $^{10}{\rm Be}(1^-) \otimes \Lambda$ & -69.27 & 8.55 &  &  & 8.84 &  &  &  &  &  &  \\
  \end{tabular}
  \end{ruledtabular}
\end{table*}
 
\subsection{Structure of core nuclei $^9$Be and $^{10}$Be}

Before the discussions on the hypernuclei, we explain the structure of the core nuclei $^{9}$Be and $^{10}$Be. 
In $^9$Be, we focus on the ground ($3/2^-$) and $1/2^+$ states. 
The $1/2^+$ state is considered to be a virtual state \cite{EPJA4.33(1999),PRC68.014310(2003)} or a resonance state \cite{PLB684.132(2010),PRC82.034001(2010)} with the well-developed $\alpha$ + $\alpha$ + $n$ cluster structure. 
By using the original parameter set of the Gogny D1S force, the excitation energy of the $1/2^+$ state $E_x (^9{\rm Be};1/2^+)$ is calculated as 2.00 MeV within the bound state approximation, whereas the central value of the observed excitation energy is $E_x^{\rm exp.} (^9{\rm Be};1/2^+) = 1.68$ MeV \cite{Ajzenberg-Selove_NPA413.1(1984)}.
For the quantitative discussions, we decreased the strength of the spin-orbit interaction of the Gogny D1S force by 5 \% to reproduce $E^{\rm exp.}_x (^9{\rm Be};1/2^+)$. 
With this modification, we obtained the excitation spectra of $^{9}$Be shown in Fig. \ref{fig:Ex.eps}(a), and the values of the total binding energy $B$ and excitation energy $E_x$ of the ground and $1/2^+$ states are summarized in Tab. \ref{Tab:table1}. 
From the proton-density distributions shown in Fig. \ref{fig:dens_i.eps} (a) and (b), it is found that the 2$\alpha$ cluster structure is more enhanced in the $1/2^+$ state compared with the ground state $3/2^-$, which leads to the difference of the nuclear quadrupole deformations $\beta$ between these states as shown in Tab. \ref{Tab:table1}. 
Furthermore, the r.m.s. radii of the $1/2^+$ state listed in Tab. \ref{Tab:table3} are larger, and thus the matter density distribution is more dilute than the ground state (see Fig. \ref{fig:dens_i.eps}).

In $^{10}$Be, we focus on the ground, $0^+_2$ and $1^-$ states. 
To reproduce the observed excitation energy of the $0^+_2$ state, $E_x^{\rm exp.} (^{10}{\rm Be};0^+_2) = 6.18$ MeV \cite{Ajzenberg-Selove_NPA413.1(1984)}, we increased the strength of the spin-orbit interaction by 17 \%, because $E_x (^{10}{\rm Be};0^+_2)$ is calculated as 4.80 MeV without this modification.
The results are shown in Fig. \ref{fig:Ex.eps}(b) and Tab. \ref{Tab:table2}. 
In the density distributions (Fig. \ref{fig:dens_i.eps}), we see the well-developed 2$\alpha$ cluster structure of the excited states $0^+_2$ and $1^-$, which is understood in terms of molecular orbit configurations around the 2$\alpha$ clusters.
In the $0^+_2$ state, it has been discussed that two valence neutrons occupy $\sigma$ orbit around the 2$\alpha$ cluster ($\sigma^2$ configuration), while the ground state $0^+_1$ has the two neutrons in $\pi$ orbit ($\pi^2$ configuration) \cite{PRC60.064304(1999)}. 
In the $1^-$ states, each of the valence neutrons occupies $\sigma$ and $\pi$ orbits ($\sigma \pi$ configuration) \cite{PRC60.064304(1999)}. 
Since $\sigma$ orbit enhances the 2$\alpha$ clustering, the inter-cluster distance increases as valence neutrons occupy $\sigma$ orbit. 
Hence, the $0^+_2$ state has the largest interval of the 2$\alpha$ clusters, while the $0^+_1$ state has the shortest. 
As a result, nuclear quadrupole deformation $\beta$ is larger in the $1^-$ and $0^+_2$ states compared with the ground state. 
The enhancement of the 2$\alpha$ clustering also leads to the difference of the r.m.s. radii. 
In Tab. \ref{Tab:table4}, it is found that the excited states $0^+_2$ and $1^-$ have the greater r.m.s. radii and their density distributions are dilute, while the ground state is compact. 
In addition to these states, three $2^+$ states also appear in the present calculation. 
Among them, the $2^+_1$ and $2^+_2$ states are classified into the $K^\pi = 0^+_1$ and $K^\pi = 0^+_2$ bands which built on the ground and $0^+_2$ states, respectively.

\subsection{Structure of $^{10}_\Lambda$Be and $^{11}_\Lambda$Be}

Let us move on the results of the hypernuclei. 
In Fig. \ref{fig:Ex.eps} (a) and (b), the excitation spectra of $^{10}_\Lambda$Be and $^{11}_\Lambda$Be calculated by using the ESC08c $\Lambda N$ interaction are also shown. 
The YN $G$-matrix interactions depend on the nuclear Fermi momentum $k_F$ due to the density dependence. We determine the value of $k_F$ by the averaged-density approximation (ADA) in the same way as Ref. \cite{PRC91.014314(2015)}, $i.e.$ the $k_F$ values in the YNG interactions are calculated from the nuclear density distributions in the ground states of $^{10}_\Lambda$Be and $^{11}_\Lambda$Be, respectively. As results, we obtained $k_F = 1.01$ fm$^{-1}$ for $^{10}_\Lambda$Be and $k_F = 1.06$ fm$^{-1}$ for $^{11}_\Lambda$Be.
With this choice of $k_F$, the $\Lambda$ binding energy $B_\Lambda$ for the ground states of $^{10}_\Lambda$Be and $^{11}_\Lambda$Be are 9.54 and 10.70 MeV, respectively. 
The $B_\Lambda$ of $^{10}_\Lambda$Be slightly overestimates the observed values, $9.11 \pm 0.22$ MeV \cite{NPA547.369(1992)} and $8.55 \pm 0.18$ MeV \cite{Gogami}. 

In $^{10}_\Lambda$Be and $^{11}_\Lambda$Be, we find that the $\Lambda$ particle brings about the significant change of the excitation spectra. 
In $^{10}_\Lambda$Be (Fig. \ref{fig:Ex.eps} (a)), it is clearly seen that the excitation energies of the positive-parity states are increased by the addition of a $\Lambda$ particle. This shift up is mainly due to the difference of the $\Lambda$ binding energies $B_\Lambda$. In Fig. \ref{fig:gain-ESC.eps}(a), we compare $B_\Lambda$ between the $^{9}{\rm Be}(3/2^-) \otimes \Lambda$ ($1^-$ and $2^-$ states) and $^{9}{\rm Be}(1/2^+) \otimes \Lambda $ ($0^+$ and $1^+$ states) doublets. 
Here, $B_\Lambda$ is calculated by using the centroid energy of each doublet. 
It is found that the $B_\Lambda$ in the $^{9}{\rm Be}(3/2^-) \otimes \Lambda$ state is larger by about 1.9 MeV than the $^{9}{\rm Be}(1/2^+) \otimes \Lambda$ state. Since the lowest threshold of $^{10}_\Lambda$Be is $^{9}_\Lambda$Be + $n$ at 3.69 MeV \cite{NPA729.337(2003),NPA639.93c(1998),Gogami}, the $^{9}{\rm Be}(1/2^+) \otimes \Lambda$ ($0^+$ and $1^+$) states are bound in spite of the large shift up. 
By the four-body cluster model calculations of $^{10}_\Lambda$Be \cite{NPA881.288(2012),PTP128.105(2012)}, the similar shift up was pointed out, in which the increase of the excitation energy was about 1.5 MeV. 

In $^{11}_\Lambda$Be, we also find the similar shift up in the excitation spectra.
In Fig. \ref{fig:Ex.eps}(b), it is seen that the excitation energies of the $K^\pi = 0^+_2$ band and negative-parity states are increased by the addition of a $\Lambda$ particle in $^{11}_\Lambda$Be. This is also due to the difference of $B_\Lambda$. 
Figure \ref{fig:gain-ESC.eps}(c) shows that $B_\Lambda$ in the $^{10}{\rm Be}(0^+_1) \otimes \Lambda $ state is larger than those in the $^{10}{\rm Be}(0^+_2) \otimes \Lambda$ and $^{10}{\rm Be}(1^-) \otimes \Lambda$ states by about 3.1 MeV and 1.9 MeV, respectively. 
As shown in Fig. \ref{fig:Ex.eps}(b), the $^{10}{\rm Be}(0^+_2) \otimes \Lambda$ state lies slightly above the lowest threshold $^{10}_\Lambda$Be + $n$ (8.70 MeV) because of the large shift up, while the $0^+_2$ state is bound in $^{10}$Be.

The difference of $B_\Lambda$ discussed above in each hypernucleus is mainly coming from the difference of the $\Lambda N$ potential energies which originates in the difference of the cluster structure of the core nuclei. 
In Tab. \ref{Tab:table1}, the expectation values of the $\Lambda N$ potential energies $V_{\Lambda N}$ are listed for the $^{9}{\rm Be}(3/2^-) \otimes \Lambda$ and $^{9}{\rm Be}(1/2^+) \otimes \Lambda$ doublets in $^{10}_\Lambda$Be. 
It is seen that $V_{\Lambda N}$ is about -16.9 MeV in the $^{9}{\rm Be}(3/2^-) \otimes \Lambda$ doublet, whereas it is about -14.4 MeV in the $^{9}{\rm Be}(1/2^+) \otimes \Lambda$ doublet. 
This is because the overlap between the $\Lambda$ and nucleons is much smaller in the $^{9}{\rm Be}(1/2^+) \otimes \Lambda$ state than that in the compact ground state $^{9}{\rm Be}(3/2^-) \otimes \Lambda$, due to the dilute density distribution of the $^{9}{\rm Be}(1/2^+) \otimes \Lambda$ state. 
In Tab. \ref{Tab:table1}, it is also seen that the $\Lambda$ kinetic energy $T_\Lambda$ is reduced in the $^{9}{\rm Be}(1/2^+) \otimes \Lambda$ doublet. This is because the distribution of the $\Lambda$ particle is more dilute in the $^{9}{\rm Be}(1/2^+) \otimes \Lambda$ state as shown in Fig. \ref{fig:dens_i.eps}(l). 
Furthermore, we see that the bindings of the nuclear part ($E_N$) are slightly shallower than the normal nucleus ($E$).
This is due to the structure change caused by a $\Lambda$ particle, which we discuss in the next section. 
Since the difference of $V_{\Lambda N}$ between the ground and the $^{9}{\rm Be}(1/2^+) \otimes \Lambda$ states is much larger than that of $T_\Lambda$ and the changes of $E_N$, the difference of $B_\Lambda$ is mainly determined by the overlap between the $\Lambda$ and nucleons of each state.

In $^{11}_\Lambda$Be, the change of the excitation spectra is understood in the same way as in $^{10}_\Lambda$Be. 
Depending on the valence neutron configuraions, the 2$\alpha$ clustering is quite different among the $^{10}{\rm Be}(0^+_1) \otimes \Lambda$, $^{10}{\rm Be}(0^+_2) \otimes \Lambda$ and $^{10}{\rm Be}(1^-) \otimes \Lambda$ states, which causes the difference of the $\Lambda N$ potential energy. 
In particular, since the density distribution of the $0^+_2$ state is dilute, the overlap between the $\Lambda$ and nucleons is much smaller. 
Therefore, the $\Lambda N$ potential energy $V_{\Lambda N}$ in the $^{10}{\rm Be}(0^+_2) \otimes \Lambda$ state is less attractive than that in the ground state $^{10}{\rm Be}(0^+_1) \otimes \Lambda$, as shown in Tab. \ref{Tab:table2}. 
In the $^{10}{\rm Be}(1^-) \otimes \Lambda$ doublet ($1/2^-$ and $3/2^-$ states), the value of $V_{\Lambda N}$ is in between the $^{10}{\rm Be}(0^+_1) \otimes \Lambda$ and $^{10}{\rm Be}(0^+_2) \otimes \Lambda$ states, because the development of the 2$\alpha$ clustering is the midst.  
In Tab. \ref{Tab:table2}, the $\Lambda$ kinetic energies $T_\Lambda$ are different corresponding to the distributions of $\Lambda$ (see also Fig. \ref{fig:dens_i.eps}(m)-(o)), whose differences are smaller than those of $V_{\Lambda N}$. 
Therefore, the difference of $B_\Lambda$ is mainly caused by that of $V_{\Lambda N}$. 
Similar behavior of $B_\Lambda$ has already been discussed for $^{12}_\Lambda$Be in our previous work \cite{PRC91.014314(2015)}. 
However, it is noted that the difference of $B_\Lambda$ between the ground and excited states in $^{11}_\Lambda$Be are much larger than that in $^{12}_\Lambda$Be. 
This is because the development of the 2$\alpha$ clustering is much different between the ground and excited states in $^{11}_\Lambda$Be compared with $^{12}_\Lambda$Be.

\subsection{Changes of the 2$\alpha$ cluster structure by a $\Lambda$ particle}

\begin{table}
  \caption{Same as Tab. \ref{Tab:table3}, but for the $0^+_1$, $0^+_2$ and $1^-$ states in $^{10}$Be and the corresponding states in $^{11}_\Lambda$Be. 
  }
  \label{Tab:table4}
  \begin{ruledtabular}
  \begin{tabular}{cccccccc}
  \multicolumn{4}{c}{$^{10}$Be} & \multicolumn{4}{c}{$^{11}_\Lambda$Be}\\
  \cline{1-4}\cline{5-8}
  $J^\pi$ & $r_p$ & $r_n$ & $r_m$ & $J^\pi$ & $r_p$ & $r_n$ & $r_m$ \\
  \hline\hline
  $0^+_1$ & 2.41 & 2.46 & 2.44 & $1/2^+_1$ & 2.36 & 2.42 & 2.39 \\
          &(2.22) &      &      &           &      &      &     \\
  $0^+_2$ & 2.92 & 3.21 & 3.10 & $1/2^+_2$ & 2.80 & 3.10 & 2.99 \\
    $1^-$ & 2.61 & 2.83 & 2.75 &   $1/2^-$ & 2.55 & 2.77 & 2.68 \\
          &      &      &      &   $3/2^-$ & 2.54 & 2.76 & 2.67 \\
  \end{tabular}
  \end{ruledtabular}
\end{table}

\begin{table}
  \caption{Intra-band $B(E2)$ values of the $K^\pi = 0^+_1$ ($2^+_1 \to  0^+_1$) and $K^\pi = 0^+_2$ ($2^+_2 \to 0^+_2$) bands in $^{10}$Be, and those in the corresponding bands of $^{11}_\Lambda$Be in unit of $e^2$fm$^4$. Numbers in parenthesis are observed value taken from Ref. \cite{NPA490.1(1988)}.}
  \label{Tab:table5}
  \begin{ruledtabular}
  \begin{tabular}{ccccc}
  \multicolumn{3}{c}{$^{10}$Be} & \multicolumn{2}{c}{$^{11}_\Lambda$Be}\\
  \cline{1-3} \cline{4-5}
  Band & Transitions & $B(E2)$ & Transitions & $B(E2)$ \\
  \hline\hline
  $K^\pi = 0^+_1$ & $2^+_1 \to 0^+_1$ &  19 & $3/2^+_1 \to 1/2^+_1$ &  8 \\
                  &                   & ($10.5 \pm 1.1$) & $5/2^+_1 \to 1/2^+_1$ &  8 \\
  $K^\pi = 0^+_2$ & $2^+_2 \to 0^+_2$ & 119 & $3/2^+_3 \to 1/2^+_2$ & 29 \\
                  &                   &     & $5/2^+_3 \to 1/2^+_2$ & 25 \\
  \end{tabular}
  \end{ruledtabular}
\end{table}

In this section, we discuss dynamical changes of the cluster structure by the addition of a $\Lambda$ particle. Particularly, we focus on the reduction of the inter-cluster distance between the 2$\alpha$ clusters and nuclear r.m.s. radii. 
In Ref. \cite{PRC91.014314(2015)}, we have found that the modification of the 2$\alpha$ clustering is rather small in $^{12}_\Lambda$Be. On the other hand, in $^{10}_\Lambda$Be, the significant reduction of the 2$\alpha$ cluster distance of the $^9{\rm Be}(1/2^+) \otimes \Lambda$ state was pointed out by Refs. \cite{NPA881.288(2012),PTP128.105(2012)}. 
Therefore, it is of interest to investigate the modification of the 2$\alpha$ cluster structure in $^{11}_\Lambda$Be and compare it with the cases of $^{10}_\Lambda$Be and $^{12}_\Lambda$Be. 

In Fig. \ref{fig:dens_i.eps}, the nuclear quadrupole deformation $\beta$ and the inter-cluster distance between the 2$\alpha$ clusters $r_{\alpha-\alpha}$ are also shown. Here, $r_{\alpha-\alpha}$ is estimated from the proton-density distributions, $i.e.$ $r_{\alpha-\alpha}$ is defined as the distance between the two peaks of the proton-density distributions corresponding to the 2$\alpha$ clusters. 
In $^{10}_\Lambda$Be, it is found that the $\beta$ and $r_{\alpha-\alpha}$ are reduced compared with those in $^{9}$Be by the attraction between the $\Lambda$ and nucleons. 
We also find that the reduction of $ r_{\alpha-\alpha}$ in the $^{9}{\rm Be}(1/2^+) \otimes \Lambda$ states (by about 7 \%) is slightly larger than that in the ground state $^{9}{\rm Be}(3/2^-) \otimes \Lambda$ (by about 5 \%), which is consistent with the four-body cluster model calculations in Refs. \cite{NPA881.288(2012),PTP128.105(2012)}. 
Corresponding to $r_{\alpha-\alpha}$, the decrease of the r.m.s. radii in the $^{9}{\rm Be}(1/2^+) \otimes \Lambda$ state is slightly larger than that in the ground state as shown in Tab. \ref{Tab:table4}. 

In $^{11}_\Lambda$Be, the degrees of structure change are clearly different between the ground ($^{10}{\rm Be}(0^+_1) \otimes \Lambda$) and excited ($^{10}{\rm Be}(0^+_2) \otimes \Lambda$ and $^{10}{\rm Be}(1^-) \otimes \Lambda$) states. 
In Fig. \ref{fig:dens_i.eps}, it is seen that the reduction of $r_{\alpha-\alpha}$ is about 8 \% in the $^{10}{\rm Be}(0^+_2) \otimes \Lambda$ and $^{10}{\rm Be}(1^-) \otimes \Lambda$ states, while the $r_{\alpha-\alpha}$ is almost unchanged in the ground state. 
Furthermore, it is also found that the reduction of the r.m.s. radii is smaller in the ground state (by about 2\% in Tab. \ref{Tab:table4}) compared with the $^{10}{\rm Be}(1^-) \otimes \Lambda$ and $^{10}{\rm Be}(0^+_2) \otimes \Lambda$ states (by about 3\%). 
These differences of the shrinkage appear as the difference of the $B(E2)$ reduction as shown in Tab. \ref{Tab:table5}. 
The intra-band $B(E2)$ values are significantly reduced in the $K^\pi = 0^+_2$ band compared with the $K^\pi = 0^+_1$ band. 
In $^{12}_\Lambda$Be \cite{PRC91.014314(2015)}, we see the similar trend of the shrinkage, in which the matter r.m.s. radius $r_m$ of the $^{11}{\rm Be}(3/2^-) \otimes \Lambda$ state with $\beta = 0.90$ is largely reduced (by 1.3\%) compared with the ground state $^{11}{\rm Be}(1/2^-) \otimes \Lambda$ with $\beta = 0.52$ (by 0.8\%). 
Since the large $\beta$ discussed above is essentially due to the development of the 2$\alpha$ clustering, it can be said that the structure change of these Be hypernuclei is dependent on the degrees of the 2$\alpha$ clustering of each state.
Finally, we also comment on the fact that the reduction of the r.m.s. radii in the $^{11}{\rm Be}(3/2^-) \otimes \Lambda$ state of $^{12}_\Lambda$Be \cite{PRC91.014314(2015)} is smaller than that in the $^{10}{\rm Be}(1^-) \otimes \Lambda$ state of $^{11}_\Lambda$Be, while their deformations are almost the same. 
We conjecture that this difference of the shrinkage between $^{11}_\Lambda$Be and $^{12}_\Lambda$Be is mainly due to the larger number of the neutrons occupying $\sigma$ orbit in the $^{11}{\rm Be}(3/2^-) \otimes \Lambda$ state (two neutrons in $\sigma$ orbits) than the $^{10}{\rm Be}(1^-) \otimes \Lambda$ state (one neutron in $\sigma$ orbit), which enhance the 2$\alpha$ clustering.

\subsection{Quantitative evaluation for the ambiguities in the calculation of $B_\Lambda$}

\begin{figure*}
  \begin{center}
    \includegraphics[keepaspectratio=true,width=172mm]{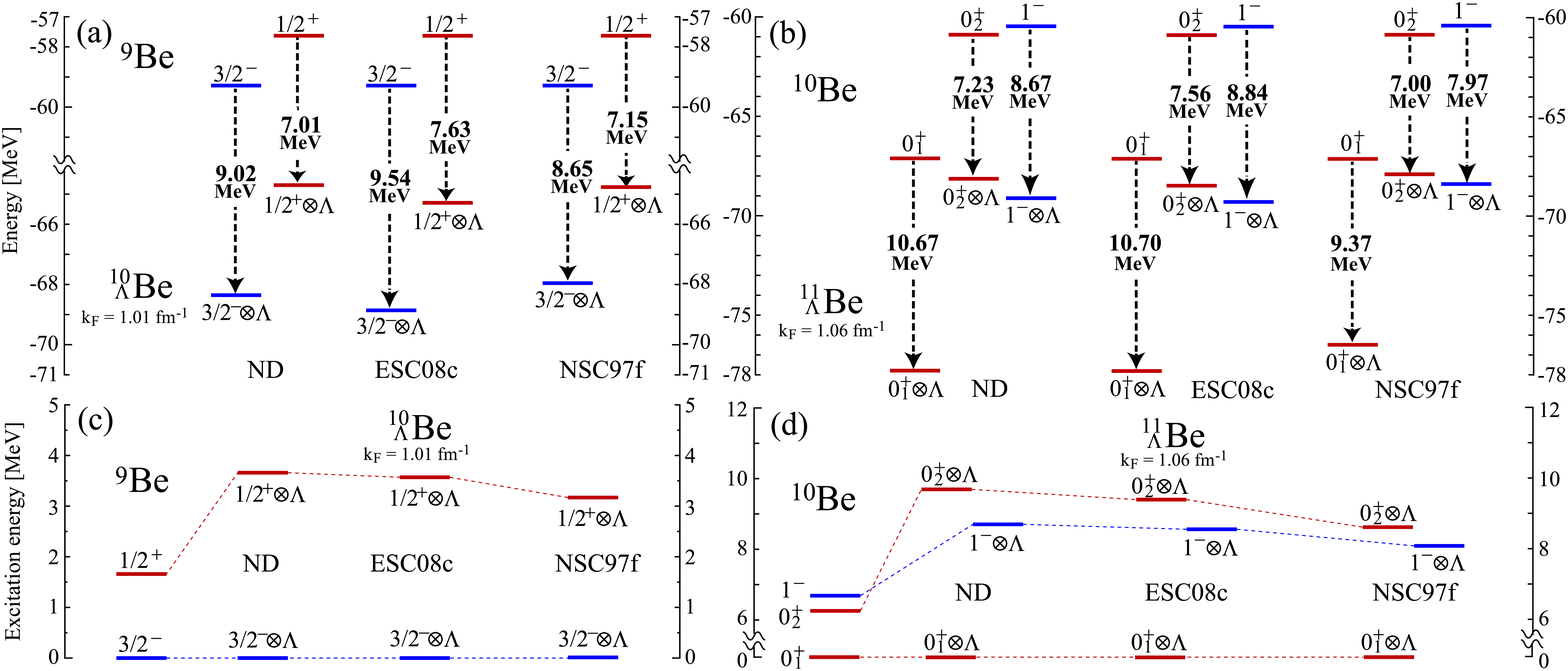}
  \end{center}
  \caption{(Color online) Comparison of $B_\Lambda$ calculated with the ND, ESC08c and NSC97f interactions in (a) $^{10}_\Lambda$Be and (b) $^{11}_\Lambda$Be. 
(c) Comparison of the excitation energies of the $^{9}{\rm Be}(1/2^+) \otimes \Lambda$ states calculated with the ND, ESC08c and NSC97f interactions. (d) Same as (c), but for the $^{10}{\rm Be}(0^+_2) \otimes \Lambda$ and $^{10}{\rm Be}(1^-) \otimes \Lambda$ states in $^{11}_\Lambda$Be.
$B_\Lambda$ in (a) and (b) and the excitation energies in (c) and (d) are calculated with the centroid energies of the doublets in each hypernucleus. }
  \label{fig:gain-Comp.eps}
\end{figure*}

In this section, we examine how the uncertainties of the $\Lambda N$ effective interactions quantitatively affect and modify the $B_\Lambda$ and excitation spectra.
First, we focus on the interaction dependence of $B_\Lambda$. 
In our previous work for $^{12}_\Lambda$Be \cite{PRC91.014314(2015)}, it has been found that the excitation energy of the $^{11}{\rm Be}(1/2^+) \otimes \Lambda$ state is dependent on the employed $\Lambda N$ interactions, $i.e.$ ND, ESC08c and NSC97f.

Figure \ref{fig:gain-Comp.eps}(a) and (b) show the comparison of $B_\Lambda$ in $^{10}_\Lambda$Be and $^{11}_\Lambda$Be among the $\Lambda N$ interactions. Here, we use the same $k_F$ in the calculations with ND and NSC97f as with ESC08c, namely $k_F = 1.01$ fm$^{-1}$ for $^{10}_\Lambda$Be and $k_F = 1.06$ fm$^{-1}$ for $^{11}_\Lambda$Be. 
In $^{10}_\Lambda$Be (Fig. \ref{fig:gain-Comp.eps}(a)), it is found that ND and NSC97f give qualitatively the same trend of $B_\Lambda$ as ESC08c in the ground and $^{9}{\rm Be}(1/2^+) \otimes \Lambda$ states. 
To evaluate the difference among the interactions quantitatively, we show the excitation energy of the $^{9}{\rm Be}(1/2^+) \otimes \Lambda$ doublet in Fig. \ref{fig:gain-Comp.eps}(c), calculated by using the centroid energies. It is seen that the difference of the excitation energies is less than 1 MeV. 
We obtain the similar result in $^{11}_\Lambda$Be.
In Fig. \ref{fig:gain-Comp.eps}(b), the trend of $B_\Lambda$ in the ground, $^{10}{\rm Be}(0^+_2) \otimes \Lambda$ and $^{10}{\rm Be}(1^-) \otimes \Lambda $ states with ND and NSC97f is almost the same as that with ESC08c. 
In Fig. \ref{fig:gain-Comp.eps}(d), it is found that the excitation energies of the $^{10}{\rm Be}(0^+_2) \otimes \Lambda$ and $^{10}{\rm Be}(1^-) \otimes \Lambda$ states differ within 1 MeV among the interactions. 

Next, we discuss the dependence of $B_\Lambda$ on the $k_F$ value on which the strength of the YNG interactions depend. 
As pointed out in Ref. \cite{PTP97.881(1997)}, the Fermi momentum $k_F$ used in the YNG interaction should be smaller in the well-developed cluster states than the compact ground states, corresponding to the lower density. 
In the discussion above, we estimated $k_F$ values by using the compact ground-state wave functions and applied them to both of the ground and exited states. Therefore, the trend of $B_\Lambda$ can be changed if the smaller values of $k_F$ are employed in the excited states. 
To investigate it, we use the different values of $k_F$ between the ground and $^{9}{\rm Be}(1/2^+) \otimes \Lambda$ states in $^{10}_\Lambda$Be. By applying the ADA treatment to each state, we obtained $k_F = 0.97$ fm$^{-1}$ in the $^{9}{\rm Be}(1/2^+) \otimes \Lambda$ state, whereas $k_F = 1.01$ fm$^{-1}$ in the ground state. With these $k_F$ values, we calculate $B_\Lambda$ and compare them in Fig. \ref{fig:gain-ESC.eps}(b). 
It is seen that the $B_\Lambda$ in the ground state is still larger than that in the $^{9}{\rm Be}(1/2^+) \otimes \Lambda $ state by about 1.5 MeV. 
In $^{11}_\Lambda$Be, we have the same conclusion as in $^{10}_\Lambda$Be. 
We also independently determine the $k_F$ values and calculate $B_\Lambda$ in the ground ($k_F = 1.06$ fm$^{-1}$), $^{10}{\rm Be}(0^+_2) \otimes \Lambda$ ($k_F = 0.98$ fm$^{-1}$), and $^{10}{\rm Be}(1^-) \otimes \Lambda$ ($k_F = 1.02$ fm$^{-1}$) states as shown in Fig. \ref{fig:gain-ESC.eps}(d). 
It confirms that the trend of $B_\Lambda$ is unchanged if the $k_F$ values are independently determined by the ADA treatment for each state. 
From Fig. \ref{fig:gain-ESC.eps}(e) and (f), it is found that the ambiguities of the excitation energies due to the $k_F$ dependence of the YNG interactions are less than 2 MeV. 
Therefore, the trend of $B_\Lambda$ in the ground and well-pronounced cluster states is unchanged if the dependence of $B_\Lambda$ on the $\Lambda N$ interaction and $k_F$ values is taken into account. 

\section{Summary} 

In this paper, we applied the HyperAMD model to $^{10}_\Lambda$Be and $^{11}_\Lambda$Be. The purpose of the present study was to reveal the changes of the excitation spectra and 2$\alpha$ cluster structure due to the addition of a $\Lambda$ particle. 
In $^{10}_\Lambda$Be, it was found that the excitation energy of the $1/2^+$ state in $^{9}$Be with well-pronounced $\alpha$ + $\alpha$ + $n$ cluster structure was largely increased. 
Despite of the large shift up, the $^{9}{\rm Be}(1/2^+) \otimes \Lambda$ state was bound in $^{10}_\Lambda$Be due to the attraction between the $\Lambda$ and nucleons.
In $^{11}_\Lambda$Be, the $0^+_2$ and $1^-$ states of $^{10}$Be were shifted up in the excitation spectra by the $\Lambda$ particle. This was due to the difference of the $\Lambda$ binding energies $B_\Lambda$, which originated in the difference of the 2$\alpha$ cluster structure. 
The changes of the excitation spectra were qualitatively the same as in $^{12}_\Lambda$Be, but quantitatively larger in $^{10}_\Lambda$Be and $^{11}_\Lambda$Be. This was because the difference of the structures between the ground and the excited state which we focused on were larger in $^{10}_\Lambda$Be and $^{11}_\Lambda$Be. 
Furthermore, we also found the significant changes of the 2$\alpha$ clustering in the excited states, in which the $\Lambda$ particle reduced the inter-cluster distance between 2$\alpha$ clusters by about 8 \%, whereas the changes of the 2$\alpha$ cluster structure were smaller in the ground states.
The difference in the structure changes of the 2$\alpha$ clustering appeared as the shrinkage of the r.m.s. radii and the reduction of the intra-band $B(E2)$ values.

\begin{acknowledgments}
One of the authors (M.K.) acknowledges support by the Grants-in-Aid for Scientific Research on Innovative Areas from MEXT (Grant No. 2404:24105008) and JSPS KAKENHI Grant 563 No. 25400240. M.I. is supported by the Special Postdoctoral Researcher Program of RIKEN. 
\end{acknowledgments}




\end{document}